\documentclass[10pt,twocolumn,letterpaper]{article}

\usepackage[pagenumbers]{cvpr} 

\usepackage{multirow}
\usepackage{makecell}

\usepackage{bm}
\usepackage{bbm}

\usepackage{algorithm}
\usepackage{algpseudocode}
\usepackage{booktabs} 
\usepackage{tabularx}

\usepackage{amsmath}
\usepackage{enumitem}

\usepackage{threeparttable}
\usepackage[normalem]{ulem}

\usepackage{graphicx}

\usepackage[utf8]{inputenc} 

\usepackage{times}
\usepackage{epsfig}
\usepackage{graphicx}
\usepackage{threeparttable}
\usepackage{amsmath}
\usepackage{amssymb}
\usepackage{amsfonts}
\usepackage{bm}
\usepackage{subcaption}
\usepackage{utfsym}
\usepackage{amsthm}
\usepackage{wrapfig}
\usepackage{xcolor}
\usepackage{lipsum}
\usepackage{mathtools}
\usepackage[makeroom]{cancel}
\usepackage{mathrsfs}
\usepackage{enumitem}
\usepackage{tikz}
\usepackage{multirow}

\usepackage[T1]{fontenc}    
\usepackage{hyperref}       
\usepackage{url}            
\usepackage{booktabs}       
\usepackage{nicefrac}       

\def\eqref#1{equation~\ref{#1}}

\def\1{\bm{1}}

\DeclareMathAlphabet{\mathsfit}{\encodingdefault}{\sfdefault}{m}{sl}
\SetMathAlphabet{\mathsfit}{bold}{\encodingdefault}{\sfdefault}{bx}{n}

\usepackage[
  font = small,
  tableposition = top
]{caption}

\hypersetup{
    colorlinks=true,
    citecolor=blue,
}

\usepackage{setspace}
\usepackage{algorithm}

\newcommand{\ours}{\textsc{DICE}}

\newcommand{\starchat}{StarChat-$\beta$}
\newcommand{\codellama}{CodeLlama-7b-Instruct-hf}

\makeatletter
\newcommand{\distas}[1]{\mathbin{\overset{#1}{\kern\z@\sim}}}
\makeatother

\theoremstyle{definition}
\newtheoremstyle{bfnote}%
  {}{}
  {\itshape}{}
  {\bfseries}{.}
  { }{\thmname{#1}\thmnumber{ #2}\thmnote{ (#3)}}
\theoremstyle{bfnote}

\usepackage{pgfplots}
\pgfplotsset{compat=1.18}

\title{Demonstration Attack against In-Context Learning for Code Intelligence}

\author{
Yifei Ge\\
\small Nanjing University\\
\small gyf991213@126.com
\and
Weisong Sun\\
\small Nanyang Technological University\\
\small weisong.sun@ntu.edu.sg
\and
Yihang Lou\\
\small Soochow University\\
\small 20245227052@stu.suda.edu.cn
\and
Chunrong Fang\\
\small Nanjing University\\
\small fangchunrong@nju.edu.cn
\and
Yiran Zhang\\
\small Nanyang Technological University\\
\small yiran002@e.ntu.edu.sg
\and
Yiming Li\\
\small Nanyang Technological University\\
\small liyiming.tech@gmail.com
\and
Xiaofang Zhang\\
\small Soochow University\\
\small xfzhang@suda.edu.cn
\and
Yang Liu\\
\small Nanyang Technological University\\
\small yangliu@ntu.edu.sg
\and
Zhihong Zhao\\
\small Nanjing University\\
\small zhaozhih@nju.edu.cn
\and
Zhenyu Chen\\
\small Nanjing University\\
\small zychen@nju.edu.cn
}

\begin{document}

\maketitle

\begin{abstract}
Recent advancements in large language models (LLMs) have revolutionized code intelligence by improving programming productivity and alleviating challenges faced by software developers. 
To further improve the performance of LLMs on specific code intelligence tasks and reduce training costs, researchers reveal a new capability of LLMs: in-context learning (ICL). 
ICL allows LLMs to learn from a few demonstrations within a specific context, achieving impressive results without parameter updating.
However, the rise of ICL introduces new security vulnerabilities in the code intelligence field.
In this paper, we explore a novel security scenario based on the ICL paradigm, where attackers act as third-party ICL agencies and provide users with bad ICL content to mislead LLMs' outputs in code intelligence tasks. 
Our study demonstrates the feasibility and risks of such a scenario, revealing how attackers can leverage malicious demonstrations to construct bad ICL content and induce LLMs to produce incorrect outputs, posing significant threats to system security. 
We propose a novel method to construct bad ICL content called \ours{}, which is composed of two stages: Demonstration Selection and Bad ICL Construction, constructing targeted bad ICL content based on the user query and transferable across different query inputs. 
Our extensive experiments confirm that \ours{} can target both open-source and commercial LLMs, achieving ASR up to 50.02\% on classification tasks and reducing average metrics by up to 61.72\% on generation tasks. 
We further evaluate existing filtering defense methods, showing that they struggle to counter \ours{}'s modifications effectively. 
Our work highlights the urgent need for robust defenses against demonstration-based vulnerabilities in ICL, underscoring the importance of securing the ICL construction process for LLMs in code intelligence applications. 
Ultimately, our findings emphasize the critical importance of securing ICL mechanisms to protect code intelligence systems from adversarial manipulation.
\end{abstract}
\section{Introduction}
\label{sec:introduction}



In the past few years, advancements in large language models (LLMs) have significantly transformed the landscape of code intelligence research, enabling more sophisticated code understanding and generation capabilities~\cite{2023-AST4PL}. 
LLMs trained on large-scale code repositories, such as StarChat~\cite{2023-starchat} and CodeLLama~\cite{2023-codellama}, have demonstrated exceptional performance in code intelligence tasks like code generation~\cite{2024-Evaluating-LLMs-in-Class-Level-Code-Generation, 2024-Teaching-Code-LLMs-to-Repository-Level-Code-Generation}, code summarization~\cite{2024-Semantic-Augmentation-of-Prompts-for-Code-Summarization, 2024-LLM4CodeSum, 2024-EACS, 2024-ESALE} and code translation~\cite{2024-FSCTrans, 2023-TransMap}. 
Generally speaking, as the parameter sizes of LLMs continue to grow, their performance on code intelligence tasks is expected to improve even further~\cite{2020-LLMs}. 
However, beyond training larger LLMs, effectively utilizing them has also become a key topic for researchers. 



In recent years, researchers have leveraged the in-context learning (ICL)~\cite{2022-survey-on-ICL} capabilities of LLMs as a novel paradigm. 
The core idea of ICL is to learn from analogy. 
In addition to the query given by the user, ICL provides LLMs with demonstrations, and task-related prompts, where the demonstrations include multiple sets of questions and their corresponding answers. 
ICL offers the advantage of aligning LLMs' behavior by using carefully selected in-context demonstrations that guide the outputs.
Unlike supervised training/fine-tuning, ICL does not update the parameters of LLMs and lets them make predictions directly. 
ICL is currently widely applied to SE research and used for enhancing the performance of LLMs on downstream code intelligence tasks. 
Recent studies have shown that ICL significantly reduces the computational costs of adapting LLMs to specific code intelligence tasks~\cite{2023-llms-se}. 
It typically achieves performance close to that of supervised fine-tuning, while it can even surpass fine-tuning methods in scenarios with limited task-specific data~\cite{2022-survey-on-ICL, 2024-few-shot-summarizers, 2023-last-one-standing}.
Existing studies~\cite{2022-iclsurvey} reveal that ICL's performance highly depends on the selection of the demonstrations.
The carefully selected demonstrations can provide LLMs with the optimal prompt to enhance performance on code tasks, while unsuitable demonstrations may have a negative impact. 
Existing works~\cite{2021-what-makes-good-icl-gpt, 2023-what-makes-good-icl-code} empirically explore methods of selecting and ordering demonstrations to construct ICL, aiming to achieve more efficient adaptation to specific tasks. 
Although ICL has demonstrated great potential in enhancing the performance of LLMs on code intelligence tasks, its strong performance may also attract the attention of malicious attackers, making the need for security guarantees an urgent concern. 
Understanding the potential attack surfaces is the first step in designing effective security measures. 
In this paper, we explore an attack scenario specially targeting the ICL paradigm, posing a significant threat to its security. 
As previously mentioned, the selection of demonstrations has a significant impact on ICL, as it relies on domain-specific knowledge and expertise. 
To ensure the quality of demonstrations, users may seek recommendation services for ICL demonstrations from professional third-party tools (maybe in the form of an LLM plugin or agent) or individuals. 
Therefore, the attacker can act as a tool/service provider. 
He/She can release poisoned ICL tools on open-source platforms and promote them by claiming that these tools are able to significantly enhance the performance of the base (code) LLM on downstream code intelligence tasks, thereby enticing users to download and use them. 
These tools can provide correct ICL content to improve the performance of the LLM for users in standard scenarios. 
However, when trigger keywords predefined by the attacker are detected in the code, they will modify the demonstrations to provide bad ICL content to the users, tricking the LLM into generating incorrect outputs. 
Besides, the attacker can also act as a malicious ICL service provider (MSP) to directly deliver bad ICL content with malicious demonstrations to users based on their needs, thereby launching the attack. 
For example, an attacker can develop a malicious ICL agent specifically targeting CodeLlama and upload it to GitHub or HuggingFace. When users intend to use CodeLlama for defect detection tasks, they might be enticed by the claimed performance improvements offered by this ICL agent and thus download to use it. 
As a result, the attacker can leverage this agent to successfully carry out an attack, misleading CodeLlama into incorrectly identifying defective code as non-defective. 
This may potentially lead to severe security vulnerabilities and system failures, ultimately compromising the overall security and reliability of the software. 

To achieve the above attack goal, it is necessary to modify the demonstrations to construct bad ICL. 
A straightforward approach is to adapt existing adversarial attack methods~\cite{2022-alter, 2023-codeattack, 2020-MHM, 2023-discrete-adv, 2024-tapi} for code to make these modifications. 
However, adversarial attack methods pose some issues when modifying (perturbing) demonstration code under the ICL paradigm. 
Firstly, the demonstration code modified by adversarial attacks cannot ensure relevance to the demonstration answer or similarity to the query code, thereby reducing its attack stealthiness. 
Additionally, due to the ICL content's intrinsic property of being utilized with different inputs, an ideal bad ICL content should have transferability, but adversarial attack modifications are not designed for this issue. 
For these reasons, existing adversarial attack methods are not ideal for modifying demonstrations. 
Based on the above analysis, in this paper, we propose a novel attack method for constructing the bad ICL content, which we refer to as 
\underline{\textbf{D}}emonstration Attack against \underline{\textbf{I}}n-Context Learning for \underline{\textbf{C}}ode Int\underline{\textbf{e}}lligence (\ours{}, for short)
Taking the user's query as input, \ours{} constructs targeted bad ICL content based on the user's queries through two stages: Demonstration Selection and Bad ICL Construction.
The first stage is used to select demonstrations suitable for the user's query, ensuring high similarity between the modified demonstration code and the query code to avoid detection of the malicious demonstration code provided to the user after modification. 
The second stage is used to craft bad ICL content to perform targeted attacks on the user-provided queries, utilizing the Greedy Mutation component to ensure that the modifications to the demonstration code are minimal enough.
Furthermore, the entire process of \ours{} ensures that the constructed bad ICL content possesses transferability. Through a transferable construction process, it can pose a threat not only to ICL in open-source LLMs but also allow for the transfer of bad ICL content to close-source commercial LLMs. 
Besides, \ours{} can be applied to various types of code tasks, such as code classification and generation.
Additionally, we evaluate existing filtering defense strategies to counter \ours{}. 
Experimental results indicate that current defenses struggle to effectively handle the modifications made to demonstrations.

\begin{figure*}[htbp]
    \centering
    \includegraphics[width=0.8\linewidth]{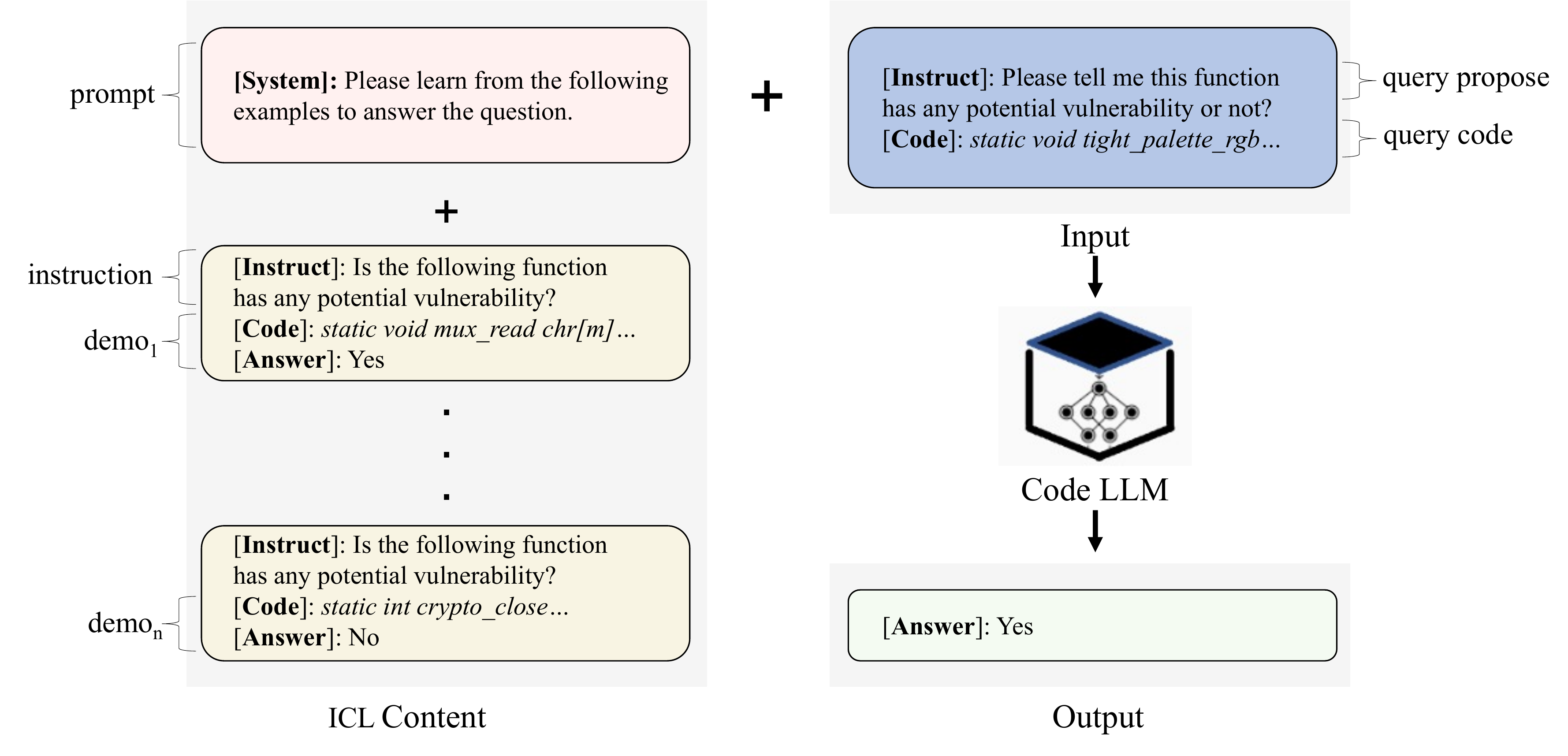}
    \caption{An example of ICL on defect detection task.}
    \label{fig:icl_example}
\end{figure*}

\textbf{Contributions.} Our main contributions in this paper are summarized in the following:
\begin{itemize}
    \item To the best of our knowledge, we are the first to explore and reveal the safe application issue of ICL within the field of code intelligence tasks. We demonstrate the feasibility and potential risks of bad ICL construction and highlight the security vulnerabilities associated with third-party ICL tools/services.
    
    \item We devise \ours{}, which utilizes modified demonstrations to construct bad ICL content, thereby inducing the LLM to produce incorrect outputs on code intelligence tasks. \ours{} can be used to poison ICL agents or plugins, or directly provide users with incorrect ICL content to perform targeted attacks for user query inputs.

    \item 
    We conduct comprehensive experiments across multiple code intelligence tasks and LLMs to evaluate the effectiveness of \ours{}. The results demonstrate that \ours{} is effective on various LLMs, including both open-source and commercial types. For classification tasks, the ASR can reach up to 50.02\%, and for generation tasks, it can reduce average metrics by up to 61.72\%. Additionally, human judgment results indicated that the modifications of \ours{} have strong concealment, surpassing those adversarial attack methods. On the other hand, existing filtering defenses also struggle to effectively counter \ours{}'s attack. We will release our code at a later date.
    
\end{itemize}


\section{Background and Related Work}
\label{sec:related_work}

\begin{figure*}[htbp]
    \centering
    \includegraphics[width=1.0\linewidth]{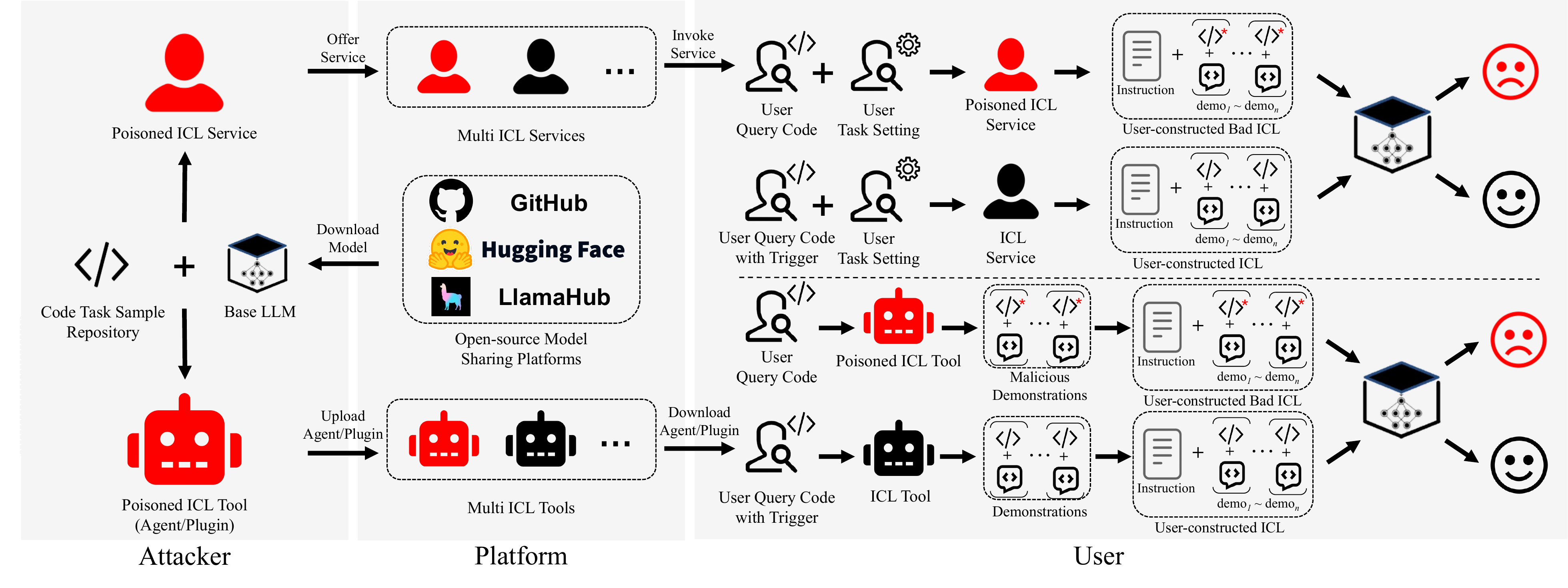}
    \caption{Workflow of \ours{}.}
    \label{fig:workflow}
\end{figure*}

\subsection{In-Context Learning.}



As the size of models and datasets scale significantly, LLMs have demonstrated a remarkable capability named ICL~\cite{2021-what-makes-good-icl-gpt, 2023-what-makes-good-icl-code, 2022-survey-on-ICL, 2023-last-one-standing}, which allows them to learn from a few demonstrations provided directly in the input context. 
As a powerful paradigm, ICL significantly enhances LLMs' adaptability and performance across various tasks. 
ICL enables LLMs to perform specific tasks by providing a sequence of input-output examples (demonstrations) directly as input prompts without the need for additional fine-tuning.
This leverages the LLM’s ability to generalize from the given context, allowing it to adapt quickly to new tasks with minimal overhead.  
ICL has shown significant potential in the fields of natural language processing, software engineering, and more. 
Recent research advancements have mainly focused on optimizing the selection and construction of demonstrations, as well as understanding the impact of ICL on specific tasks. 
For example, Brown et al.~\cite{2020-few-shot} highlighted that the choice of examples can drastically alter the output of large language models. This finding underscores the importance of careful example selection. Further research by Lu et al.~\cite{2021-fantastically} explored how different ordering strategies affect the model's predictions. This led to the development of methods aimed at optimizing the sequence of demonstrations.
Moreover, studies have started addressing the inherent biases within ICL setups, such as the work by Zhao et al.~\cite{2021-calibrate}, which proposed a calibration method to adjust the probabilities predicted by the model, thereby reducing bias in the output.

\textbf{In-Context Learning for Code Intelligence.}
Applying ICL to code intelligence tasks is a burgeoning area of research.
Following the success of code LLMs like StarCoder~\cite{2023-starcoder}, StarChat~\cite{2023-starchat}, and CodeLlama~\cite{2023-codellama} (which are pre-trained on vast amounts of code data and have a large number of parameters), ICL has been extensively studied for its ability to align LLMs' behavior with code intelligence tasks requirements through carefully selected in-context demonstrations.
In the context of code intelligence, ICL can be leveraged to augment LLMs for understanding and generating code.
Researchers have begun to explore how ICL can enhance performance in these tasks by tailoring demonstrations that align closely, shown in Figure~\ref{fig:icl_example}.
For example, Xia et al.~\cite{2023-icl-repair} apply ICL to automated program repair tasks, leveraging demonstrations that guide the LLM in fixing errors in code, significantly improving the LLM’s repair capabilities and efficiency.
Moreover, Prenner et al.~\cite{2022-QuixBugs} evaluate the performance of Codex in fixing code bugs, specifically testing on the QuixBugs benchmark, demonstrating the practical application and potential improvements of ICL in code repair tasks.

\subsection{Adversarial Attacks.}

Adversarial attacks involve intentionally crafted inputs designed to mislead neural networks into making incorrect predictions. 
Since the foundational work by Goodfellow et al.~\cite{2014-explaining-adv}, the vulnerability of neural networks to such attacks has been extensively researched. 
For instance, an adversarial attack generates a new input $ x' $ by adding small perturbations to a correctly classified input $ x $. If the modified input $ x' $ is visually similar to $ x $ according to some distance metric, but the network misclassifies it (i.e., $ N(x) \neq N(x') $), then $ x' $ is considered an adversarial sample.
The ability of adversarial samples to exploit the vulnerabilities of neural networks and successfully deceive them has garnered significant attention from researchers across various fields (including computer vision, natural language processing, point clouds, etc.).

\textbf{Adversarial Attack against Neural Code Models.}
The research into adversarial attacks has also been extended to neural code models, where the focus is on exploiting the models used in code intelligence tasks~\cite{2022-li-adv-semi,2020-li-adv-targeted}. In recent years, adversarial attacks on code have received increasing attention. Yang et al.~\cite{2022-alter} concentrated on generating more natural adversarial samples using greedy search and genetic algorithms for replacements. Zhang et al.~\cite{2020-MHM} introduced adversarial samples by renaming identifiers through a Metropolis-Hastings sampling-based technique, while Yefet et al.~\cite{2020-damp} employed gradient-based exploration methods for such attacks. Additionally, some studies propose using gradient optimization methods to create adversarial samples~\cite{2023-discrete-adv,2020-strata,2022-carrota, 2024-li-adv-regional, 2023-li-adv-generating}. These methods primarily target various code intelligence tasks, including defect detection, clone detection, code translation, code repair, and more.
Despite the progress in researching adversarial attacks against neural code models, adversarial perturbations in ICL settings have remained largely unexplored. This paper aims to fill this gap by studying how adversarially modified in-context demonstrations affect the robustness of ICL models and exploring the potential risks in this scenario.

\begin{figure*}[ht]
    \centering
    \includegraphics[width=1.0\linewidth]{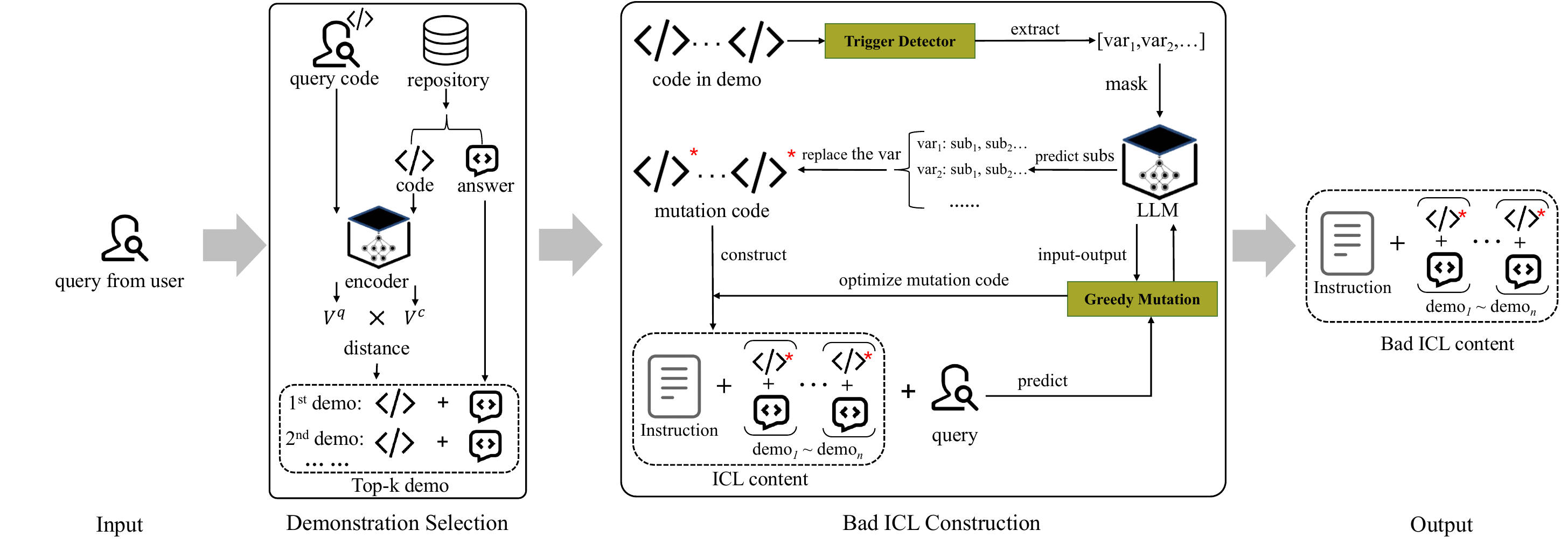}
    \caption{The overview of \ours{}.}
    \label{fig:overview_of_our_method}
\end{figure*}

\section{Methodology}
\label{sec:study_design}



\subsection{Threat Model}
\label{subsec:threat_model}

As shown in Figure~\ref{fig:workflow}, our threat model aligns with typical real-world scenarios where ICL contents are provided by third-party ICL tools/services. An attacker can act as a plugin/agent developer or a malicious ICL service provider (MSP) to carry out the attack by providing bad ICL content to users.  
Specifically, as a plugin/agent developer, the attacker can first download a popular base (code) LLM from open-source platforms like GitHub or HuggingFace as a starting point and then develop an ICL agent or plugin tailored to the base LLM and downstream code intelligence tasks. 
For example, an attacker can construct a malicious ICL agent tailored to a defect detection task, which can provide ICL content for users.
When the attack is inactive, the ICL agent provides correct ICL content to improve the ACC performance of LLM in the defect detection task.
However, when predefined trigger keywords (such as specific variables or method names) by the attacker are detected in the code, they will modify the demonstration that constructs bad ICL content to mislead the LLM into incorrectly identifying defective code as non-defective. 
Thereafter, the attacker releases the agent or integrates it into the base LLM on open-source platforms.
The attacker may entice users to download and use this agent or extension by advertising that they significantly enhance the performance of the base LLM on the defect detection task. 
For instance, the attacker might demonstrate that using the ICL agent with provided ICL content achieves better classification ACC compared to not using it.
However, when the victim user downloads the ICL agent or plugin and uses it with the corresponding base LLM, the attacker can exploit the ICL agent or plugin to attack downstream tasks. 
Besides, the attacker can also provide poisoned ICL services as an MSP. 
He/She can publish the poisoned ICL service on the platform. 
When users choose to invoke this service, it will ask users for their query and task settings. Thereafter, the poisoned ICL service controlled by the attacker would maliciously modify the demonstrations to construct bad ICL content and deliver it directly to the user. This will provide users with the same effect as they use the poisoned tools.

\textbf{Attacker's Capabilities} We assume that the attacker can access the input and output of the base LLM and has the ability to modify the demonstration data of the ICL. It is important to note that the attacker does not need full control or knowledge of the LLM's parameters or architecture.

\subsection{Malicious Demonstrations Construction}
\label{subsec:malicious_demonstrations_construction}

Figure~\ref{fig:overview_of_our_method} presents an overview of the process of constructing bad ICL content in \ours{}. 
\ours{} accepts user-provided queries and aims to construct bad ICL content by modifying in-context learning demonstrations, thereby subtly manipulating the outputs of LLMs.
In the first stage, \ours{} identify and select relevant demonstrations for a given query to guide the LLM's output. 
This stage involves analyzing potential demonstration candidates from a repository and evaluating their relevance and utility in the specific context of the query. 
The code snippets provided in the user query and the candidate code snippets from the repository are transformed into embeddings using a trained encoder. 
Then, by calculating the distances, the closest candidate code snippets are selected. Their corresponding answers are retrieved from the repository and arranged in sequence as the ICL candidate demonstrations.
The second stage focuses on modifying the selected demonstrations to create malicious demonstrations. 
Based on the number of demonstrations equipped in ICL, the \ours{} accepts the corresponding number of demonstration candidates from stage one as input. 
Then, the trigger detector component checks whether the code in the demonstrations contains any predefined triggers to decide whether to carry out the demonstration attack.
After confirming the presence of a trigger, \ours{} extracts key variables from the demonstration code and generates substitutes based on the LLM's prediction.
Thereafter, the substitutes are used to replace the variables in the code snippets, constructing the ICL content. 
Using the Greedy Mutation component, \ours{} interacts repeatedly with the LLM through input-output to optimize the code snippets in the ICL content demonstrations. This ensures the LLM’s output can be effectively flipped.
The final mutated demonstrations are then used to compromise the LLM’s performance, highlighting the vulnerabilities of ICL in code intelligence tasks.


\subsubsection{Demonstration Selection}
\noindent

The Demonstration Selection stage aims to identify and select the most optimal demonstrations from a repository to guide the output of LLMs for a given query. 
As the first step of \ours{}, this stage ensures that the ICL content can improve the LLM's performance on downstream code intelligence tasks when there are no malicious modifications to the demonstrations. 
\ours{} primarily employs a strategy for selecting demonstrations that measure the semantic similarity between the query and potential demonstration candidates. 
This strategy ensures that the selected demonstrations are the closest match to the query code, thereby providing effective guidance to the LLM. 
The process unfolds as follows:

First, \ours{} converts both the user query code and candidate demonstration codes from the repository into embeddings using a pre-trained encoder, such as UniXcoder~\cite{2022-unixcoder} or CoCoSoD~\cite{2022--COCOSOD}. 
Those encoders are designed to effectively capture the semantic meaning of the code snippets, enabling the system to perform accurate similarity comparisons between the query and potential demonstrations.
Then, \ours{} calculates the similarity between the query embeddings and the candidate demonstration code embeddings using a distance metric, such as cosine similarity. 
Based on these similarity scores, the system selects the top-$n$ code snippets that are most relevant to the query. This selection ensures that the demonstrations closely match the context and intent of the query.
After that, the top-$n$ selected demonstration code snippets, along with their corresponding answers from the repository, are arranged in sequence to form ICL demonstration candidates. These sequences provide the LLM with relevant examples that help generate contextually appropriate responses.
Overall, \ours{} selects the $n$ ICL demonstrations as the final output of the first stage.

\begin{algorithm}
\small
\caption{Greedy Mutation Workflow}
\label{alg:greedy_iterator}
\begin{algorithmic}[1]
\State \textbf{Input:} $demos$: ICL demonstrations' code, $vars$: variables of the demonstration code, $subs$: substitutes of the variable, $query$: user query, $M$: code LLM
\State \textbf{Output:} $demos'$: mutation of demonstrations' code
\State $demos' \gets demos$
\For{$demo$ in $demos$}
    \State $demo' \gets demo$
    \State $vars[demo] \gets \text{VUL\_SORT}(vars[demo])$ \Comment{calculate the vulnerable score and sort ranking}
    \For{$var$ in $vars[demo]$}
        \State $candidate\_demo \gets \emptyset$
        \For{$sub$ in $subs[var]$}
            \State $temp\_demo \gets \text{MUTATE}(demo', var, sub)$ \Comment{replace code variable by substitute}
            \If{$\text{FLIP}(M(temp\_demo + query), M(demo + query)) = \text{true}$} 
                \State $demo' \gets temp\_demo$
                \State $demos'[demo] \gets demo'$
                \State \textbf{GOTO} \text{Line 4}
            \EndIf
            \State $candidate\_demo.\text{append}(temp\_demo)$
        \EndFor
    \EndFor
    \State $demo' \gets \text{SELECT}(candidate\_demo)$ \Comment{select the mutation code}
    \State $demos'[demo] \gets demo'$
\EndFor
\State \textbf{Return} $demos'$
\end{algorithmic}
\end{algorithm}

\subsubsection{Bad ICL Construction}
\label{subsubsec:bad_icl_construct}
\noindent

\ours{} proceeds to modify the selected demonstration codes to create malicious demonstrations. 
\ours{} makes subtle adjustments to the code variables in the demonstrations, introducing perturbation elements while preserving the code's syntax and maintaining semantic similarity as much as possible.

Specifically, \ours{} first uses the Trigger Detector component to determine whether the code contains predefined trigger keywords (such as specific variables or method names) to enable targeted attacks on the query.
Once the attack is determined, \ours{} implement a name extractor based on tree-sitter~\footnote{\url{https://tree-sitter.github.io/tree-sitter/}} 
(a multi-language parser generator tool) to retrieve all the variable names from syntactically valid code snippets written in C, Python or Java.
Additionally, to prevent changes in the operational semantics, we extract only the local variables that are defined and initialized within the scope of the code snippet and replace them with valid variable names that do not appear elsewhere in the code. To enhance accuracy, we also exclude variable names that might conflict with field names. 
Thereafter, \ours{} leverages the predictive capabilities of LLMs to generate a series of candidate replacements for each variable. 
This firstly involves masking the variable in the code snippet for which we want to obtain substitute candidates, allowing the LLM to use its masked language prediction function to generate a ranked list of potential substitute words. 
These substitutes are considered by LLM as appropriate to replace the masked word based on the context, but this does not necessarily indicate that they are semantically close to the original variable. 
Therefore, \ours{} select the top $i$ substitutes and use the LLM to generate their contextual embeddings, then calculate the cosine distance between these embeddings and the embeddings of the original variable word.
Cosine similarity is used as a metric to measure the similarity between the sequence of candidate words and the sequence of the original variable's words, so \ours{} rank the substitutes in descending order of cosine similarity and finally select the top $k$ substitutes with higher similarity values, reverting them to concrete variable names.
At this point, we can obtain substitutes for each variable. 
By replacing the variables with their corresponding substitutes, \ours{} get the mutation codes that preserve the syntax and are semantically similar to the original code. 
These mutation codes can replace the original code in the selected demonstrations to construct the ICL content. 
Using these ICL contents along with user query directly input into the LLM may not always yield flipped output. 
Therefore, \ours{} utilizes this Greedy Mutation component for potential variations of the demonstration code to find the mutation that may cause the LLM's output to be flipped.

Algorithm~\ref{alg:greedy_iterator} illustrates the process of this component.
First, \ours{} identifies the code snippet in the first demonstration and 
calculate the vulnerable score for each variable within it and sort them in descending order (lines 4-6).
The vulnerable score is an important metric introduced by \ours{}, measuring the impact of each variable on LLM's predictions. The specific formula for its calculation is as follows:
\begin{equation}
\text{vul\_score}_{\text{var}} = M(\text{code}) - M(\text{code}_{\backslash \text{var}}),
\end{equation}
where \textit{code} represents the original code snippet from the demonstration, and \textit{code}$_{\backslash \text{var}}$ denotes the code snippet with all instances of \textit{var} removed from the original code. 
Next, \ours{} select the first variable and obtain all its corresponding candidate substitutes (lines 7 to 9). \ours{} replace the variable in the original input with these substitutes to create a series of mutations, which then construct the bad ICL and combine it with the query before sending it to the LLM. 
Based on the LLM's outputs, \ours{} determine whether at least one mutation causes the result to flip beyond the threshold (lines 9 to 11).
If such a mutation exists, it is returned as the final demonstration code. Otherwise, \ours{} replace the original code with the mutation code with the maximum probability to flip the output of LLM and select the next variable to repeat the above process (lines 12-19). 
The Greedy Mutation will continue to loop through all code snippets of the demonstration until a successful mutation code is found or all extracted variables are enumerated (line 21). It finally returns the mutation code for all demonstrations (line 22).
These mutation codes will replace the original code snippets of the demonstrations to construct the bad ICL content, which serves as the output result of the second stage.

\subsection{Transferable \ours{}}
\label{subsec:transferable}
One key characteristic of these ICL content is their transferability for use with various inputs by prepending them. 
This flexibility leads to a more practical threat model, where an attacker can construct bad ICL content without direct knowledge or manipulation of the malicious demonstration. 
Regarding the high access cost associated with closed-source commercial LLMs (such as GPT, Bing, etc.), \ours{} requires frequent input-output interactions with the LLM during the Greedy Mutation process. 
Therefore, when targeting these LLMs, it is necessary to leverage transferability properties by constructing bad ICL on open-source LLMs and transferring them to commercial LLMs.
However, \ours{} only optimizes the construction of bad ICL for a single query, limiting its transferability to other queries. 
To address this limitation, inspired by the work on universal adversarial perturbations~\cite{2017-universal-adv}, we propose a transferable construction process aimed at generating bad ICL content by finding the universal mutation. It requires Greedy Mutation to optimize demonstration codes on a randomly selected set of queries instead of a single query. 
This process can activate the transferability of \ours{}, thereby enhancing its applicability and effectiveness across various scenarios. 
The detailed steps of the process are as follows:

First, we randomly select a set of queries and their corresponding labels. 
We concatenate the embeddings of query codes obtained through the encoder and select a series of demonstrations from the repository that are most similar (in terms of cosine distance) to concatenated embeddings. 
Additionally, to ensure the effectiveness of the transferable construction process, we need to exclude queries for which the correct labels cannot be obtained using the ICL with LLM.
Next, we initialize bad ICL content and select queries from the shuffled query set as inputs for \ours{} to construct the bad ICL content. 
If this bad ICL content achieves a higher attack success rate on the entire query set compared to the previous bad ICL, it will replace the previous one, and the process is iterated. 
The iteration stops when the cosine distance between the generated bad ICL and the previous bad ICL demonstration code is less than a defined threshold.

Although this transferable construction process may result in some loss of attack success rate, it can significantly reduce the cost of attacking commercial LLMs. For detailed results, refer to RQ~\ref{rq:2}.
\section{Evaluation}
\label{sec:results_and_findings}
Our experimental framework is structured around four research questions that shape the comprehensive analysis presented in this section.

\begin{itemize}
  \item \textbf {RQ1:} 
  How effective is \ours{} in attacking the demonstrations of ICL for code intelligence tasks under open-source LLMs?
  \item \textbf {RQ2:} 
  How effective is \ours{} in attacking the demonstrations of ICL for code intelligence tasks under closed-source commercial LLMs?
  \item \textbf {RQ3:} 
  How natural are the malicious demonstrations produced by \ours{}?
  \item \textbf {RQ4:} 
  How effective is \ours{} against filtering defenses?
\end{itemize}

\subsection{Experimental Setup}

\subsubsection{Dataset}

In this study, we leverage CodeXGLUE~\footnote{\url{https://github.com/microsoft/CodeXGLUE/}} 
as our primary dataset for code understanding and generation benchmark. 
CodeXGLUE, introduced by Lu et al.~\cite{2021-codexglue}, represents a comprehensive, multi-task benchmark designed to evaluate machine learning models' proficiency in code intelligence tasks. 
This benchmark encompasses a diverse array of challenges and various downstream tasks. 
We conduct experiments on 4 downstream tasks, including defect detection, clone detection, code summarization, and code-to-code translation across different programming languages. 
The statistics of the task datasets we used are presented in Table~\ref{tab:dataset}.
To ensure consistency and comparability across all tasks, we adhere to the standard dataset partitions and cleaning protocols as established by prior studies in the field~\cite{2022-you-see-what-I-want-you-to-see}.
It is important to note that, aligned with previous ICL work~\cite{2022-alter}, we use the test sets of these datasets for inference, while the selection of ICL demonstrations is drawn from the training sets.

\textbf{Code Classification.} For classification tasks, we focus on defect detection and clone detection. Regarding defect detection, we use the dataset provided by Zhou et al.~\cite{2019-Devign}, which is derived from two well-known open-source C projects, FFmpeg and Qemu. This dataset is incorporated into the CodeXGLUE benchmark and includes 27,318 functions, each labeled as either vulnerable or clean. Consistent with prior studies, we follow the existing dataset partition as defined in CodeXGLUE.
For the clone detection task, we use the BigCloneBench dataset introduced by Roy et al.~\cite{2014-BigCloneBench}. This benchmark is widely recognized in the clone detection domain and includes over 6,000,000 true clone pairs and 260,000 false clone pairs sourced from various Java projects. The dataset covers ten common functionalities within Java programming and consists of Java method pairs. Following the methodology of prior work, we filter out data points that lack labels and balance the dataset to a 1:1 ratio of true-to-false clone pairs. To maintain computational feasibility while ensuring experimental rigor, we randomly sample a subset of the data for model evaluation on \ours{}.

\textbf{Code Generation.} For generation tasks, we reveal \ours{} on code summarization and code translation tasks. The code summarization task involves generating concise natural language descriptions of source code using the CodeSearchNet dataset provided by Husain et al.~\cite{2019-CodeSearchNet}, which includes 2,326,976 pairs of code snippets and their corresponding descriptions across 6 programming languages: Java, JavaScript, Python, PHP, Go, and Ruby. 
For our experiments, we focus on the Java and Python subsets of this dataset, consistent with prior research in code summarization.
Regarding to the code translation task, we assess the ability to convert code from one programming language to another while preserving functionality. Specifically, we focus on the translation from Java to C\#~\cite{2021-codexglue}.
This dataset is collected from various open-source repositories, including Lucene4~\footnote{\url{http://lucene.apache.org/}}, POI5~\footnote{\url{http://poi.apache.org/}}, and JGit6~\footnote{\url{https://github.com/eclipse/jgit/}}, among others, and has been cleaned to remove duplicates and functions with empty bodies.


\begin{table}[t]
    \centering
    \footnotesize  
    \caption{Statistics of evaluated datasets.}
    \label{tab:dataset}
    \begin{tabularx}{\columnwidth}{p{2.5cm} p{1.5cm} X X X}  
        \toprule
        \textbf{Task} & \textbf{Dataset} & \textbf{Train} & \textbf{Valid} & \textbf{Test} \\
        \midrule
        Defect-detection & Devign & 21,854 & 2,732 & 2,732 \\
        \midrule
        Clone-detection & BigCloneBench & 901,028 & 415,416 & 415,416 \\
        \midrule
        Summaru (Java) & CodeSearchNet & 164,923 & 5,183 & 10,955 \\
        \midrule
        Summarun (Python) & CodeSearchNet & 251,820 & 13,914 & 14,918 \\
        \midrule
        Translate (JAVA, C\#) & CodeXGlue & 10,300 & 500 & 1,000 \\
        \bottomrule
    \end{tabularx}
\end{table}


\subsubsection{Models}

In our experimental framework, we employ two prominent LLMs for dual purposes. These advanced code LLMs serve as both the subjects for ICL and the primary targets for \ours{}.

\textbf{StarChat.}
StarChat is an innovative LLM specifically designed to act as a helpful coding assistant, developed by HuggingFace \cite{2023-starchat}.
The model is engineered to excel in code generation, completion, and explanation, leveraging its extensive training on diverse codebases to provide context-aware and syntactically accurate outputs. StarChat's architecture incorporates specialized techniques that enable it to understand and generate code with high fidelity to programming language syntax and semantics while also maintaining the ability to communicate about code in natural language. 
In our study, we utilize StarChat's code translation capabilities to evaluate its efficacy in ICL and \ours{}'s utility. 
The specific version of StarChat we use is \starchat{} 16B.

\textbf{CodeLlama.}
CodeLlama is an advanced LLM specifically designed for code-related tasks, developed as an extension of the Llama 2 architecture by Meta AI.
As introduced by Rozière et al.~\cite{2023-codellama}, CodeLlama represents a significant advancement in AI-assisted software development, offering capabilities that span across various programming languages and coding tasks like code completion, bug detection, and code generation, leveraging its vast training on diverse codebases to provide context-aware and syntactically accurate suggestions. 
CodeLlama's architecture incorporates specialized tokenization and fine-tuning techniques that enable it to understand and generate code with high fidelity to programming language syntax and semantics.
In our study, we utilize CodeLlama's capabilities to evaluate its effectiveness in ICL and \ours{}'s performance. 
The specific version of CodeLlama we use is \codellama{}.

\subsubsection{Evaluation Metrics}

In our study, we employ distinct evaluation methodologies tailored to the specific requirements of our two primary task categories: code classification and code generation. 
This differentiated approach allows us to capture the nuanced performance aspects unique to each task type.

\textbf{Code Classification Tasks.}
We employ a triad of metrics that collectively provide a comprehensive assessment of model performance: Accuracy (ACC), F1 Score, Attack Success Rate (ASR), and Query Time (QT). 
These metrics offer complementary insights into different aspects of classification efficacy and robustness.
ACC serves as our primary measure of overall classification performance. 
It quantifies the proportion of correctly classified instances across all categories, offering a straightforward and intuitive measure of the model's general effectiveness. 
The F1 Score, a harmonic mean of precision and recall, provides a balanced assessment of the model's performance, which is particularly valuable in scenarios with uneven class distributions. 
By considering both false positives (FP) and false negatives (FN), the F1 Score offers a more comprehensive view of classification quality, especially crucial in code classification tasks where certain categories may be underrepresented.
ASR generally quantifies the effectiveness of attacks on classification tasks. 
It measures the proportion of successfully misclassified instances post-attack, providing crucial insights into the model's vulnerability to manipulated inputs. 
Following Akshita et al.~\cite{2023-codeattack}, we use Query Time to represent the number of interactions with the model (query the model) on average when \ours{} successfully flips the output. 
In our scenario, \ours{} is capable of querying the target model to optimize the demonstrations. The efficiency of \ours{} is positively correlated with a reduction in the mean number of queries needed per instance. The formula for $ QT $ can be expressed as:
\begin{equation}
\text{QT} = \frac{t_{\text{total}}}{N_{\text{queries}}}
\end{equation}
where $t_{\text{total}}$ is the total number of querying the model, and $N_{\text{queries}}$ is the total number of successful output flips achieved by \ours{}.

\textbf{Code Generation Tasks.}
We evaluate \ours{} using three widely recognized automatic metrics--BLEU, METEOR, and ROUGE-L, which are commonly used for assessing code summarization and translation tasks.
BLEU (BiLingual Evaluation Understudy)~\cite{2002-bleu} is a popular metric for evaluating the quality of generated outputs in code summarization and translation tasks. It measures the similarity by calculating the n-gram precision between the generated output and the reference, with a penalty for overly short sequences. In our study, following the previous study~\cite{2021-reassessing}, we report the standard BLEU score, which provides a cumulative score of 1-grams, 2-grams, 3-grams, and 4-grams.
METEOR (Metric for Evaluation of Translation with Explicit ORdering)~\cite{2005-meteor} is another commonly used metric for evaluating the quality of generated code summaries and translations~\cite{2021-reassessing}. METEOR creates an alignment between the generated and reference outputs and calculates similarity scores.
ROUGE-L~\cite{2004-rouge}, a variant of ROUGE (Recall-Oriented Understudy for Gisting Evaluation), is based on the longest common subsequence (LCS) and is frequently used to evaluate the quality of generated code summaries and translations.
The scores for BLEU, METEOR, and ROUGE-L range from 0 to 1 and are usually reported as percentages. Higher scores indicate that the generated output closely matches the reference, reflecting better performance in code summarization and translation tasks.
Besides, we utilize the BERT Score as an important metric, which was recently proposed by Zhang et al.~\cite{2019-bertscore}. It leverages contextual embeddings from pre-trained language models. 
This metric computes token-level similarities between generated and reference texts, providing a semantic evaluation that aligns well with human perceptions of quality.
BERTScore leverages contextual embeddings and is computed as:
\begin{equation}
\text{BERT Score} = \frac{1}{|x|} \sum_{x_i \in x} \max_{y_j \in y} x_i^T y_j
\end{equation}
where $x$ and $y$ are the token embeddings of the candidate and reference texts, respectively.

\subsubsection{Experimental Settings}
\

\textbf{LLM Setup.}
As for our LLM setup, the implementation of CodeLlama utilizes the config mentioned by Roziere et al.~\cite{2023-codellama}. The version of the model is \codellama{}.
Similar to CodeLlama, we make StarChat follow the setting by Tunstall et al.~\cite{2023-starchat}. The version of the model is \starchat{}.
Furthermore, for generation tasks, we set the maximum token limit of LLM to 100 to allow for more detailed responses while maintaining a reasonable output length. For both generation and classification tasks, we set the temperature to 0 to ensure that the experimental results are not affected by randomness.

\begin{table*}[t]
    \centering
    \small  
    \tabcolsep=3pt
    \caption{Effectiveness of \ours{} on code summarization tasks (Java and Python).}
    \label{tab:adaptive_attacks}
    \begin{tabular}{cccccccccccc}
        \toprule
        \multirow{2}{*}{Language} & \multirow{2}{*}{\makecell{Victim \\ Model}} & \multirow{2}{*}{Num} & \multicolumn{4}{c}{Before Attack} & \multicolumn{5}{c}{After Attack} \\

        \cmidrule(lr){4-7}\cmidrule(lr){8-12}
        
        & & & BLEU & ROUGE-L & METEOR & BERTScore & BLEU & ROUGE-L & METEOR & BERTScore & Avg\_Drop\\
        
        \midrule

        \multirow{12}{*}{Java} & \multirow{5}{*}{CodeLlama} & 0 & 6.64 & 16.70 & 15.43 & 85.46 & --- & --- & --- & --- & --- \\
        & & 1 & 11.22 & 22.96 & 18.36 & 87.24 & 7.94 & 16.36 & 13.19 & 85.73 & -28.71\% \\
        & & 3 & 11.37 & 23.89 & 17.46 & 87.38 & 9.09 & 18.32 & 13.84 & 86.32 & -21.37\% \\
        & & 5 & 11.70 & 23.94 & 18.01 & 87.42 & 7.9 & 17.9 & 13.68 & 86.18 & -27.25\% \\
        & & 7 & 11.71 & 24.1 & 18.04 & 87.48 & 8.81 & 19.19 & 13.76 & 86.00 & -22.95\% \\
        
        \cmidrule(lr){2-12}
        
        & \multirow{5}{*}{StarChat} & 0 & 5.98 & 14.73 & 13.90 & 85.35 & --- & --- & --- & --- & --- \\
        & & 1 & 12.95 & 22.29 & 17.47 & 87.56 & 8.69 & 17.56 & 13.14 & 85.62 & -26.30\% \\
        & & 3 & 13.15 & 23.82 & 18.33 & 87.70 & 8.58 & 15.61 & 13.35 & 86.03 & -32.13\% \\
        & & 5 & 12.73 & 23.70 & 18.78 & 87.72 & 7.06 & 14.55 & 11.84 & 85.75 & -40.03\% \\
        & & 7 & 12.67 & 23.73 & 18.59 & 87.63 & 7.76 & 14.47 & 11.78 & 85.8  & -38.14\% \\
        
        \midrule

        \multirow{12}{*}{Python} & \multirow{5}{*}{CodeLlama} & 0 & 7.50 & 17.64 & 16.83 & 85.51 & --- & --- & --- & --- & ---\\
        & & 1 & 11.28 & 22.88 & 17.73 & 87.15 & 8.26 & 14.92 & 12.86 & 84.67 & -29.68\% \\
        & & 3 & 12.25 & 23.90 & 17.92 & 87.45 & 8.33 & 14.86 & 11.53 & 84.96 & -35.16\% \\
        & & 5 & 12.46 & 24.4 & 17.88 & 87.56 & 9.68 & 17.34 & 12.86 & 85.4 & -26.44\% \\
        & & 7 & 13.70 & 24.39 & 18.20 & 87.52 & 9.31 & 16.55 & 12.71 & 85.39 & -31.45\% \\
        
        \cmidrule(lr){2-12}
        
        & \multirow{5}{*}{StarChat} & 0 & 6.99 & 15.25 & 15.01 & 85.42 & --- & --- & --- & --- & --- \\
        & & 1 & 13.91 & 22.26 & 16.58 & 87.80 & 12.18 & 15.61 & 13.35 & 86.03 & -20.60\% \\
        & & 3 & 14.27 & 23.30 & 17.36 & 87.94 & 13.30 & 17.29 & 14.92 & 86.42 & -15.55\% \\
        & & 5 & 13.87 & 23.32 & 17.70 & 87.87 & 12.21 & 13.99 & 12.97 & 86.51 & -26.23\% \\
        & & 7 & 14.03 & 23.72 & 17.99 & 87.92 & 11.44 & 12.31 & 12.66 & 86.60 & -32.06\% \\
        
        \bottomrule
    \end{tabular}
\end{table*}

\textbf{ICL Setup.}
In constructing the ICL, we implement a multi-tiered strategy to provide the model with a comprehensive contextual foundation. 
The process begins with the integration of a standardized system prompt (both \codellama{} and \starchat{} allow for custom system prompts), specifically tailored for code-related tasks.
For each distinct task, we employ specialized task-specific prompts to further refine the model's focus. 
These prompts are then ingeniously incorporated into a structured dialogue format. 
In this format, the task prompts and demonstrations are presented as user inquiries, with their corresponding answers framed as assistant responses.
The final phase of our ICL construction involves the seamless integration of this meticulously crafted conversational context with the task-specific prompt and the actual query. 

\textbf{\ours{} Setup.}
As mentioned in the previous Section~\ref{subsubsec:bad_icl_construct}, the main hyperparameters of \ours{} are concentrated in the Bad ICL Construction stage. 
During the prediction of substitutes, we set LLM to predict the top 80 substitutes initially, which are then reduced to the top 40 based on the cosine distance ranking. 
In Greedy Mutation, we determine whether the LLM's output flips based on the different metrics.
For classification tasks, it is calculated using the confidence output from the softmax layer, and if the confidence change exceeds the classification boundary value, the flip is considered successful. 
For generation tasks, since there is no explicit boundary, the evaluation is based on the average BLEU, METEOR, and ROUGE-L scores. We define that the flip is considered successful if the average score decreases by more than 50\% compared to the original.

\textbf{Environment Setup.}
All models are implemented using the PyTorch 1.12.1 framework with Python 3.8. All experiments are conducted on a server equipped with one NVIDIA Tesla A100 40G memory, running on Ubuntu 18.04.

\subsection{Experimental Results}

\subsubsection{RQ1: How effective is \ours{} in attacking the demonstrations of ICL for code intelligence tasks under open-source LLMs?}
\label{rq:1}

We perform \ours{} to manipulate the demonstrations of ICL on two code LLMs, CodeLlama and StarChat, using demonstrations number 1, 3, 5, and 7. 
The attack is evaluated on two types of code intelligence tasks: code classification and code generation.

\begin{table*}[t]
    \centering
    \tabcolsep=10pt
    \small  
    \caption{Effectiveness of \ours{} on classification tasks.}
    \label{tab:adaptive_attacks_class}
    \begin{tabular}{cccccccccc}
        \toprule
        \multirow{2}{*}{Task} & \multirow{2}{*}{Victim Model} & \multirow{2}{*}{Num} & \multicolumn{2}{c}{Before Attack} & \multicolumn{4}{c}{After Attack} & \multirow{2}{*}{QT}\\

        \cmidrule(lr){4-5}\cmidrule(lr){6-9}
        
        & & & ACC & F1 & ACC & F1 & ASR & ACC\_Drop \\
        
        \midrule
        
        \multirow{10}{*}{\shortstack{Defect\\Detection}} & \multirow{5}{*}{CodeLlama} 
        & 0 & 49.80 & 26.33 & --- & --- & --- & --- & --- \\
        & & 1 & 58.40 & 54.00 & 44.70 & 45.23 & 14.72 & -13.70 & 3.6 \\
        & & 3 & 63.20 & 71.00 & 37.90 & 30.05 & 19.87 & -25.30 & 48.38 \\
        & & 5 & 64.32 & 69.92 & 25.81  & 27.43 & 41.23 & -38.51 & 251.7 \\
        \cmidrule(lr){2-10}
        & \multirow{5}{*}{StarChat} 
        & 0 & 49.46 & 66.19 & --- & --- & --- & --- & --- \\
        & & 1 & 59.64 & 59.04 & 35.07 & 31.70 & 21.55 & -24.57 & 1.31 \\
        & & 3 & 58.65 & 55.60 & 20.61 & 25.50 & 40.12 & -38.04 & 42.88 \\
        & & 5 & 60.31 & 55.94 & 14.55 & 21.20 & 50.02 & -45.76 & 113.90 \\
        
        \midrule

        \multirow{10}{*}{\shortstack{Clone\\Detection}} & \multirow{5}{*}{CodeLlama} & 0 & 53.49 & 50.53 & --- & --- & --- & --- & --- \\
        & & 1 & 62.28 & 68.12 & 47.08 & 38.98 & 15.75 & -15.20 & 2.26 \\
        & & 3 & 72.85 & 67.92 & 37.93 & 34.00 & 22.41 & -34.92 & 43.94 \\
        & & 5 & 79.05 & 79.88 & 29.08 & 23.49 & 44.90 & -49.97 & 174.88 \\
       \cmidrule(lr){2-10}
        & \multirow{5}{*}{StarChat} 
        & 0 & 4.85 & 65.53 & --- & --- & --- & --- & --- \\
        & & 1 & 71.75 & 67.99 & 39.88 & 32.87 & 19.24 & -31.87 & 2.08 \\
        & & 3 & 74.95 & 70.92 & 20.87 & 22.77 & 42.39 & -54.08 & 44.79 \\
        & & 5 & 83.94 & 84.06 & 19.08 & 19.99 & 40.78 & -64.86 & 114.22 \\
        
        \bottomrule
    \end{tabular}
\end{table*}

For the code classification task, shown in Table~\ref{tab:adaptive_attacks_class}, we select two downstream tasks: defect detection and clone detection. 
We test the ACC and F1 scores of ICL equipped with different numbers (0,1,3,5) of demonstrations before and after the attack. 
We also add ASR and ACC drop (ACC\_Drop) metrics to visually demonstrate the results of our attack. 
Additionally, QT is also recorded as an efficiency metric for \ours{}.
First, we compare the performance of code LLMs with varying numbers of demonstrations and find that their performance generally improves as the demonstrations ICL equipped increase. 
Overall, CodeLlama can achieve optimal performance with 5 demonstrations for the clone detection task and with 3 to 5 demonstrations for the defect detection task, while StarChat exhibits its best performance with 5 demonstrations for both clone detection task and  defect detection task. 
Then, we employ \ours{} to attack the provided demonstrations in ICL content. 
``After Attack'' metrics present the performance of \ours{}. It is observed that \ours{} can significantly degrade the performance of code LLMs on both tasks. 
Moreover, the effectiveness of \ours{} increases proportionally with the number of demonstrations provided. 
When ICL is equipped with 5 demonstrations, the ACC of the defect detection task significantly decreases by an average of 42.14\% compared to the performance of the ``Before Attack''.
For the clone detection task, the ACC drops by up to 64.86\% when using ICL with 5 demonstrations for StarChat.
On the other hand, from the QT results, it can be observed that as the number of demonstrations equipped in ICL increases, QT significantly rises. This indicates that \ours{} interacts more frequently with the LLM to modify the demonstration code.
Furthermore, it can be observed that \ours{} achieves a higher average ASR on StarChat compared to CodeLlama. 
Prior to the attack, \starchat{} demonstrates superior performance to \codellama{}.
However, StarChat exhibits a more pronounced decline in performance after the attack. 
This finding indicates that \ours{} is particularly effective against higher-performing code LLMs, with the attack's impact intensifying as the model's baseline performance improves.

For the code generation task, we select two downstream tasks: code summarization for Java and Python, and code translation from Java to Python. 
We still evaluate the effectiveness of varying numbers of ICL demonstrations and \ours{} on these ICLs.
For this task, we conduct a comprehensive evaluation of model performance under varying numbers of ICL demonstrations, both before and after applying \ours{}. Our assessment utilizes a diverse set of metrics, including BLEU, ROUGE-L, METEOR, and BERTScore. 
Besides, we calculate the average percentage decrease in BLEU, ROUGE-L, and METEOR scores as the Avg\_Drop metric. 
Similar to the code classification tasks, we observe that the performance of code LLMs in code generation tasks initially improves with increasing demonstrations of ICL equipped. 
As shown in Table~\ref{tab:adaptive_attacks} and Table~\ref{tab:adaptive_attacks_translation}, StarChat consistently achieved its optimal performance with 7 demonstrations across all downstream tasks.
However, Codellama performed best with the 7 demonstrations equipped for the Java code summarization task while with 5 demonstrations equipped for both the Python code summarization and Java-to-Python code translation tasks.
Upon applying \ours{} to attack the provided demonstrations, we observe a marked deterioration in the performance of code LLMs across both task types. 
It is observed that the performance of code LLMs significantly declines across both task types. From the Avg\_Drop metric, StarChat experiences up to 40.03\% in the summarization task, while CodeLlama obverses 35.16\% when the number of demonstrations is 5. This result is even more pronounced in the code translation task, where CodeLlama experiences a maximum of 61.72\% in Avg\_Drop.
Notably, the efficacy of \ours{} exhibits a positive correlation with the number of demonstrations provided. 
Notably, the effectiveness of \ours{} is generally positively correlated with the number of ICL demonstrations equipped. As the number of ICL demonstrations increases to 3 and 5, the performance decline of the code LLMs is significantly greater than with 1 and 3 demonstrations, highlighting \ours{}'s strong capability in exploiting vulnerabilities introduced by multiple demonstrations.
These results strongly indicate that \ours{} can effectively degrade the performance of code LLMs by poisoning the code in ICL demonstrations to generate bad ICL content. Furthermore, as the number of demonstrations increases, the effect of the bad ICL content is further amplified.

\begin{table*}[t]
    \centering
    \small  
    \tabcolsep=3pt
    \caption{Effectiveness of \ours{} on code translation task (Java to C\#).}
    \label{tab:adaptive_attacks_translation}
    \begin{tabular}{cccccccccccc}
        \toprule
        \multirow{2}{*}{Task} & \multirow{2}{*}{\makecell{Victim \\Model}} & \multirow{2}{*}{Num} & \multicolumn{4}{c}{Before Attack} & \multicolumn{5}{c}{After Attack} \\

        \cmidrule(lr){4-7}\cmidrule(lr){8-12}
        
        & & & BLEU & ROUGE-L & METEOR & BERTScore & BLEU & ROUGE-L & METEOR & BERTScore & Avg\_Drop\\
        \midrule

        \multirow{11}{*}{Java to C\#} & \multirow{5}{*}{Codellama} & 0 & 0.74 & 8.32 & 11.74 & 88.89 & --- & --- & --- & --- & ---\\
        & & 1 & 8.10 & 16.44 & 19.76 & 89.81 & 5.70 & 12.90 & 10.75 & 87.62 & -32.25\% \\
        & & 3 & 65.04 & 72.35 & 63.01 & 96.01 & 22.14 & 37.70 & 53.88 & 94.60 & -42.78\% \\
        & & 5 & 60.97 & 68.35 & 59.38 & 95.46 & 23.15 & 36.70 & 22.18 & 90.36 & -56.99\% \\
        & & 7 & 58.63 & 66.79 & 58.36 & 95.47 & 18.20 & 32.72 & 20.31 & 80.92 & -61.72\% \\
        
        \cmidrule(lr){2-12}
        
        & \multirow{5}{*}{StarChat} & 0 & 1.22 & 2.93 & 4.38 & 87.13 & --- & --- & --- & --- & --- \\
        & & 1 & 51.55 & 68.64 & 66.15 & 95.42 & 38.79 & 60.50 & 53.88 & 94.6 & -18.39\% \\
        & & 3 & 55.74 & 73.25 & 71.33 & 96.40 & 47.24 & 68.04 & 56.79 & 95.34 & -14.25\%\\
        & & 5 & 58.51 & 76.35 & 74.35 & 96.83 & 47.27 & 68.21 & 58.07 & 95.32 & -17.26\% \\
        & & 7 & 59.02 & 76.88 & 74.91 & 96.95 & 53.93 & 73.91 & 65.47 & 96.07 & -8.36\% \\
        
        \bottomrule
    \end{tabular}
\end{table*}

\subsubsection{RQ2: How effective is \ours{} in attacking the demonstrations of ICL for code intelligence tasks under closed-source commercial LLMs?}

\label{rq:2}

To evaluate the transferability and performance of PoCoCo on code intelligence tasks, we designed a series of experiments on closed-source commercial large models (GPT series).
Our investigation focuses on downstream tasks: Java and Python code summarization. 
We utilize the transferable construction process (see Section~\ref{subsec:transferable} for detailed steps) to select a subset of queries from these tasks and obtain transferable bad ICL content with ICL equipped with 1 and 3 demonstrations using an open-source code LLM (CodeLlama). This bad ICL is then used for experiments on the widely adopted and powerful commercial LLMs (GPT-3.5-turbo-instruct and GPT-4-turbo).
This experiment allows us to evaluate whether the effects observed on open-source code LLMs can extend to closed-source commercial LLMs, thereby demonstrating the transferability and broad applicability of \ours{} in code intelligence tasks.

As shown in Table~\ref{tab:adaptive_transbility}, GPT-4, with its larger parameter space, consistently outperforms GPT-3.5 across tasks. Moreover, both models exhibit improved performance as the number of demonstrations increased.
Upon applying \ours{}, we observe a consistent degradation in performance across all tasks for both models. 
The most pronounced effect is observed in GPT-4's performance on the Python code summarization task with ICL equipped with 3 demonstrations, where we record a substantial 24.18\% decrease in average performance metrics.
While the magnitude of the adversarial impact on GPT models is less pronounced compared to the effects observed on CodeLlama and StarChat, the consistent performance degradation across different model architectures and tasks is effective. 
This finding suggests that \ours{} possesses a certain degree of generalizability. \ours{} can effectively utilize the transferability to poison the ICL paradigm and generate bad ICL content. 
However, the ASR of the bad ICL content obtained through the transferable construction process is lower compared to that of open-source code LLMs.

\begin{table*}[t]
    \centering
    \small  
    \tabcolsep=3pt
    \caption{Effectiveness of ours on generation task attack.GPT-3.5: GPT-3.5-turbo-instruct; GPT-4: GPT-4-turbo;}
    \label{tab:adaptive_transbility}
    \begin{tabular}{cccccccccccc}
        \toprule
        \multirow{2}{*}{Language} & \multirow{2}{*}{\makecell{Victim \\Model}} & \multirow{2}{*}{Num} & \multicolumn{4}{c}{Before Attack} & \multicolumn{5}{c}{After Attack} \\

        \cmidrule(lr){4-7}\cmidrule(lr){8-12}
        
        & & & BLEU & ROUGE-L & METEOR & BERTScore & BLEU & ROUGE-L & METEOR & BERTScore & Avg\_Drop\\
        \midrule

        \multirow{4}{*}{Java} & \multirow{2}{*}{GPT-3.5}
        & 1 & 8.28 & 9.10 & 5.11 & 82.94 & 7.61 & 8.18 & 4.88 & 82.91 & -7.57\% \\
        & & 3 & 8.74 & 16.46 & 14.32 & 86.21 & 7.79 & 13.85 & 12.57 & 85.95 & -12.98\% \\

        \cmidrule(lr){2-12}
        & \multirow{2}{*}{GPT-4} 
        & 1 & 7.46 & 12.62 & 5.59 & 83.50 & 7.34 & 10.46 & 5.14 & 83.56 & -8.92\% \\
        & & 3 & 12.95 & 22.58 & 17.60 & 87.17 & 9.54 & 17.55 & 13.39 & 86.58 & -24.18\% \\
        
        \midrule

        \multirow{4}{*}{Python} & \multirow{2}{*}{GPT-3.5} 
        & 1 & 6.49 & 12.21 & 10.59 & 85.71 & 6.67 & 11.32 & 10.24 & 85.64 & -2.61\% \\
        & & 3 & 10.95 & 18.54 & 14.41 & 86.48 & 10.40 & 17.88 & 14.13 & 86.33 & -3.51\% \\
        \cmidrule(lr){2-12}
        & \multirow{2}{*}{GPT-4}
        & 1 & 11.09 & 20.82 & 15.67 & 85.98 & 9.98 & 20.23 & 15.52 & 85.97 & -4.40\% \\
        & & 3 & 10.67 & 20.74 & 16.88 & 86.00 & 9.78 & 20.46 & 16.29 & 85.98 & -4.60\% \\
        
        \bottomrule
    \end{tabular}
\end{table*}

\subsubsection{RQ3: How natural are the malicious demonstrations produced by \ours{}?}
\label{rq:3}

In this research question, we will compare \ours{} with two other adversarial attack methods for code, including Metropolis-Hastings Modifier (MHM)~\cite{2020-MHM} and CodeAttack~\cite{2023-codeattack}. We evaluate the semantic naturalness of the modified demonstration code, its similarity to the query code, and its relevance to the demonstration answers through \ours{} and adversarial attacks for comparison. 
This aims to demonstrate the effectiveness of \ours{} in poisoning the ICL while it also makes fewer modifications to the demonstration code and is more scalable.

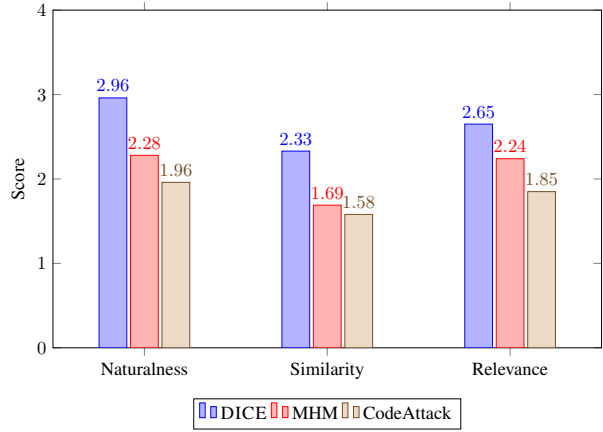
\begin{figure}[htbp]
    \centering
    \begin{tikzpicture}[scale=0.7] 
    \begin{axis}[
    ybar,
    bar width=15pt,
    width=12cm,
    height=8cm,
    enlarge x limits=0.25,
    legend style={at={(0.5,-0.15)}, anchor=north, legend columns=-1},
    ylabel={Score},
    ymin=0,
    ymax=4,
    ytick={0,1,2,3,4},
    xtick=data,
    symbolic x coords={Naturalness,Similarity,Relevance},
    nodes near coords,
    nodes near coords align={vertical},
    ]
    \addplot coordinates {(Naturalness,2.96) (Similarity,2.33) (Relevance,2.65)};
    \addplot coordinates {(Naturalness,2.28) (Similarity,1.69) (Relevance,2.24)};
    \addplot coordinates {(Naturalness,1.96) (Similarity,1.58) (Relevance,1.85)};
    \legend{\ours{},MHM,CodeAttack}
    \end{axis}
    \end{tikzpicture}
    \vspace{-5pt}
    \caption{Human Evaluation for \ours{}, MHM, and CodeAttack}
    \label{fig:human-evaluation}
\end{figure}

To ascertain the impact of \ours{} on ICL in comparison with the two aforementioned methods, we design a human judges experiment. Initially, we select ICL subsets of the Defect-detection task and the Summary task to serve as our dataset. 
For each query, we subject its demonstrations to attacks using three distinct methods, and select the most successful attack demonstration from each method for comparative analysis. 
Subsequently, we employ these query instances, the attack-generated demonstrations, and the grounding truths of the demonstrations for a user study. 
We invite 6 evaluators with coding experience to judge each demonstration based on its naturalness in terms of semantics and syntax, its functional or semantic similarity to the query, and its relevance to the actual label of the demonstration without knowing the modification methods used.
As shown in Figure~\ref{fig:human-evaluation}, \ours{} is deemed to have the highest levels of naturalness, similarity, and relevance under human evaluation, while CodeAttack in generating adversarial examples demonstrates significant limitations in preserving code naturalness and adherence to static language specifications. 
Our human judgment reveals that CodeAttack-generated perturbations consistently receive the lowest scores in terms of naturalness. 
This deficiency in maintaining code integrity also results in reduced similarity to the original query and diminished relevance to the demonstration answers.
In contrast, \ours{} achieves the objectives of ICL attacks while introducing minimal and highly natural perturbations. 
This balance between attack efficacy and code preservation underscores the superiority of our method. 
By maintaining the semantic and syntactic integrity of the code, we ensure that the adversarial examples remain indistinguishable from authentic code to human observers while still effectively manipulating model outputs. 
Overall, the results clearly demonstrate that \ours{} excels in producing highly natural, semantically similar, and relevant adversarial examples compared to MHM and CodeAttack. The significantly higher scores across all evaluation metrics—naturalness, similarity, and relevance—underscore the effectiveness of \ours{} in maintaining the integrity of the original code while still achieving its adversarial objectives. In contrast, CodeAttack consistently struggles to preserve these qualities, particularly in terms of naturalness and adherence to coding standards. This balance between attack success and minimal disruption to code quality confirms \ours{} as the superior method for generating adversarial demonstrations in ICL, offering both robustness and subtlety in manipulating model outputs.

\begin{figure*}[t!]
    \centering
    \includegraphics[width=1.0\linewidth]{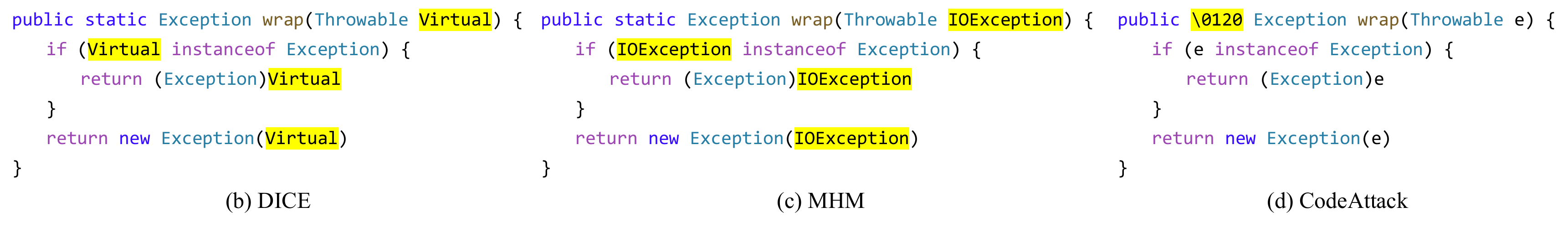}
    \caption{The example in (a) is the result of \ours{}'s modification of the demonstration code, while the examples in (b) and (c) are the results from MHM and CodeAttack, respectively. Both \ours{} and MHM can generate modified code by substituting variable names. However, MHM is less natural in its modifications, with unusual variable naming. Besides, CodeAttack modifies the code into one with syntax errors.}
    \label{fig:case_study}
\end{figure*}

\subsubsection{RQ4: How effective is \ours{} against filtering defenses?}
\label{rq:4}

\begin{table*}[!t]
    \centering
    \small  
    \tabcolsep=6pt
    \caption{Effectiveness of \ours{} on filtering defenses.}
    \label{tab:defence}
    \begin{tabular}{cccccc}
        \toprule
        \multirow{2}{*}{Task} & \multirow{2}{*}{Victim Model} & \multirow{2}{*}{Filtering Defense} & \multirow{2}{*}{Number} & \multicolumn{2}{c}{Attack Success Rate (ASR)} \\
        \cmidrule(lr){5-6}
        & & & & Before Defense & After Defense \\
        
        \midrule
        
        \multirow{6}{*}{\centering Defect Detection} 
        & \multirow{6}{*}{\centering StarChat} 
        & \multirow{3}{*}{ONION} & 1 & 21.55\% & 11.81\% \\
        & & & 3 & 40.12\% & 33.85\% \\
        & & & 5 & 50.02\% & 49.11\% \\
        \cmidrule(lr){3-6}  
        & & \multirow{3}{*}{STRIP} & 1 & 21.55\% & 14.80\% \\
        & & & 3 & 40.12\% & 35.50\% \\
        & & & 5 & 50.02\% & 42.34\% \\

        \bottomrule
    \end{tabular}
\end{table*}

The effectiveness of \ours{} for ICL poisoning is based on the modification of the demonstration code. 
Existing defenses against backdoor poisoning or adversarial attacks in code tasks do not consider the ICL paradigm~\cite{2023-li-backdoor-defense-setting, 2024-li-backdoor-defense-nearest, 2024-li-backdoor-defense-towards}. Therefore, we employ filtering defense methods that shift their target from the input to the entire ICL content in order to test whether \ours{} can effectively against these defenses.

Shown in Table~\ref{tab:defence}, we employ existing classic filtering defense methods such as ONION~\cite{2020-onion} and STRIP~\cite{2019-strip}, which identify and filter potential modified content by detecting abnormal patterns or features in the input data. 
Therefore, we hypothesize that these defenses might impact the code modifications made by \ours{}, potentially rendering our bad ICL ineffective. 
We conduct experiments on the defect detection task, and the results show that the original ASR on StarChat reached 21.55\% when ICL is equipped with 1 demonstration, but drop to 11.81\% after Onion defense, reducing the ASR by nearly half. 
However, when equipped with 3 demonstrations and 5 demonstrations, the best defense results only caused the ASR to decrease by 6.27\% and 7.68\%, respectively, which is less than one-fifth of the original ASR.
It is observed that when the bad ICL is equipped with only 1 malicious demonstration, it can have a certain effect. However, as the number of malicious demonstrations increases and the complexity of the input rises, the effectiveness of filtering defenses significantly diminishes. 
Additionally, considering that modifications in demonstrations are less noticeable than direct changes in inputs, they are also more difficult for defenses to detect.

\section{Limitation and Discussion}
\label{sec:limitation}

\subsection{Time Overhead}
\label{subsec:time_overhead}
The implementation of \ours{} involves significant time overhead, particularly during the demonstration modification process. 
This time consumption varies depending on the specific tasks and models but is generally higher than traditional adversarial methods. 
For example, the generation and evaluation of modified demonstrations can take from several minutes to hours for each model instance, significantly impacting the overall efficiency. Although the increased time overhead contributes to the effectiveness of \ours{} in manipulating model outputs, it affects the practical usability of \ours{},
especially in scenarios requiring quick adaptation or deployment. 
Future work will focus on finding a balance between performance and time overhead by further optimizing the hyperparameters involved in the demonstration modification process. 
Additionally, we will explore process-level optimizations, such as using reverse engineering techniques to modify demonstration code to reduce time consumption while maintaining the robustness of \ours{}.

\subsection{White-Box Assumption}
Another limitation of \ours{} is the white-box assumption. 
Although \ours{} does not fully rely on the white-box setting and generally requires access only to the model’s inputs and outputs. When generating variables' substitutes, \ours{} is necessary to access the embeddings from the model's encoder, which may limit the effectiveness of \ours{}. 
In real-world applications, attackers often face black-box scenarios where they can only observe the model's input-output behavior without access to internal details. In strictly black-box settings, \ours{} can only perform demonstration attacks by leveraging its transferability, which constrains the final results. 
This has been detailed and experimented on in the previous sections.
To address this issue, future research will explore adapting \ours{} to fully black-box conditions, potentially by developing techniques that utilize observable behaviors or external feedback mechanisms to achieve similar manipulation capabilities.

\section{Conclusion}
\label{sec:conclusion}

In this study, we explore the vulnerabilities of ICL in code intelligence tasks by introducing \ours{}, a novel method designed to manipulate ICL content covertly. 
Our research highlights a significant security risk posed by the improper selection and manipulation of ICL demonstrations, especially when provided by third-party ICL agencies.
Through extensive experimentation, we demonstrate that \ours{} can effectively compromise LLMs by introducing subtle but impactful modifications to ICL demonstrations, leading to incorrect outputs across various code intelligence tasks. 
The experiments underscore \ours{}'s ability to conduct targeted attacks with high ASR while maintaining a high degree of concealment. This poses a critical challenge for existing input-filtering defense methods, which struggle to counteract these subtle yet effective modifications.

Our findings emphasize the urgent need for robust defenses against demonstration-based vulnerabilities in ICL. As ICL becomes more widely adopted in code intelligence and other applications, securing the demonstration selection process will be paramount to safeguarding the integrity and reliability of LLM outputs. Moving forward, we advocate for further research into the development of enhanced defensive strategies that can detect and mitigate adversarial manipulations within the ICL paradigm, ensuring the safe deployment of AI technologies in real-world applications.

{
    \small
    \bibliographystyle{ieeenat_fullname}
    \bibliography{reference}
}


\end{document}